# Limit to mass sensitivity of nanoresonators with random rough surfaces due to intrinsic sources and interactions with the surrounding gas

G. Palasantzas [a]

Zernike Institute of Advanced Materials, University of Groningen, Nijneborgh 4, 9747 AG Groningen, The Netherlands

**Abstract**

We investigate initially the influence of thermomechanical and momentum exchange noise on the limit to mass sensitivity $\Delta m$ of nanoresonators with random rough surfaces, which are characterized by the roughness amplitude $w$, the correlation length $\xi$, and the roughness exponent $0<H<1$. In fact, $\Delta m$ increases with increasing roughness (decreasing $H$ and/or increasing ratio $w/\xi$) if the quality factor associated with thermomechanical noise is larger than that due to momentum exchange noise. Finally, the influence of adsorption-desorption noise, which is also influenced by the surface morphology, is shown to play minimal role in presence of the other two noise sources.

*Pacs numbers:* 85.85.+j, 73.50.Td, 68.55.-a, 74.62.Fj

[a] Corresponding author: G.Palasantzas@rug.nl

Nanoelectromechanical systems (NEMS) are class of devices, which combine the advantages of mechanical systems, e.g., applicability as sensor systems and robustness to electrical shocks, with the speed and large scale integration of silicon electronics [1-5]. Moreover, nanomechanical structures provide extremely high resonance frequencies, minuscule active masses and very small force constants. An additional important attribute is their relatively high quality factors *Q* ($\sim10^3$-$10^6$) [3-5]. These functionalities translate into diverse possibilities for high mass sensitivity at high resonance frequencies. In general the operation of resonant mass sensors is based on relating a frequency shift that is proportional to the inertial mass of deposited molecules.

The resonator sensitivity is determined by the effective vibratory mass of the resonator (which is determined by geometry, configuration and material properties of the resonant structure), and the stability of the device resonance frequency [5]. The frequency stability is governed by extrinsic processes (originating from the transducer and readout circuitry) [6, 7], and intrinsic processes fundamental to the nanomechanical resonator itself [3-5, 8]. The enhanced sensitivity that is attainable in NEMS [5, 9], in combination with ultrasensitive transduction techniques [3-5, 10], indicates that fundamental fluctuation processes are likely to determine their overall sensing performance.

Furthermore, studies of SiC/Si NEMS have shown that devices operational in the UHF/microwave regime had low surface roughness, while devices with rougher surfaces could not be operated higher than the VHF regime [11]. Also studies of Si nanowires have shown the quality factor to decrease by an increment of the surface area to volume ratio [12]. Recently random surface roughness was shown to affect the quality factor and the dynamic range of nanoresonators [13]. As an overall outcome we can state that the previous studies showed that surface effects play a dominant role in NEMS. These considerations motivate the present work to explore how fluctuation processes impose ultimate limits on the sensitivity of nanosize inertial mass sensors by taking into account the morphology of their surfaces in presence of thermomechanical and momentum exchange noise. Indeed, both types of noise lead to displacement fluctuations.

Thermomechanical noise arises from coupling between a mechanical resonator and its dissipative reservoir. This coupling damps the driven motion of the resonator and induces spatial fluctuations in the resonator's position peaking at the mechanical resonance frequencies [14, 15]. They could be a dominant source of frequency noise at a given mode of vibration, thereby setting the ultimate limits of detection for a dynamic micromechanical sensor [14, 16]. Notably, due to its small heat capacity, a nanoresonator can also be subject to large temperature fluctuations inducing frequency fluctuations since dimensions and material parameters depend on temperature [3, 4]. Furthermore, the resonator can undergo gas damping due to impingement and momentum exchange of gas molecules on its surface [3-5], as well as mass loading due to molecule adsorption-desorption [3-5, 13].

For thermomechanical noise the spectral density of frequency fluctuations is given by [3-5] $S_\omega(\omega)_{th} = (\omega_o^5 K_B T / E_c Q_{in}^3)[(\omega^2 - \omega_0^2)^2 + (\omega\omega_0)^2 / Q_{in}^2]^{-1}$ with $Q_{in}$ the intrinsic quality factor of the resonator, and $E_C = M_{eff} \omega_o^2 <u_c^2>$ is the maximum drive energy when the resonator is driven at a constant mean square amplitude $<u_c>$ by a voltage-controlled oscillator [4]. For momentum exchange noise, the noise spectral density is given by $S_\omega(\omega)_{m-e} = (\omega_o^5 K_B T / E_c Q_{gas}^3)[(\omega^2 - \omega_0^2)^2 + (\omega\omega_0)^2 / Q_{gas}^2]^{-1}$ [3, 6, 11] assuming that the resonator operates with quality factor $Q_{gas}$ in the molecular regime. This corresponds to molecule mean free path $L_{mph}$ ($= 0.23 K_B T / P d^2$; for a dilute gas of pressure $P$ assuming the molecules as hard spheres with diameter $d$) [17] larger than the beam width $w_b$ (<0.1L and $L$ the beam length) or equivalently large Knudsen numbers $K_n = L_{mph} / w_b > 10$ [17]. Moreover, we have $Q_{gas} = M_{eff} \omega_o \sqrt{K_B T / m} (P A_{rou})^{-1}$ with $m$ the molecule mass, $M_{eff}$ the effective resonator mass that oscillates, and $A_{rou}$ rough surface area of the resonator [13].

In presence of both types of noise, the total quality factor Q is given by $1/Q = 1/Q_{in} + 1/Q_{gas,r}$, and the corresponding spectral density by $S_\omega(\omega) = (\omega_o^5 K_B T / E_c Q^3)[(\omega^2 - \omega_0^2)^2 + (\omega\omega_0)^2/Q^2]^{-1}$. Therefore, the frequency fluctuations yield a frequency shift $\delta\omega$ and an associated limit to mass sensitivity $\Delta m$ [3-5], which are given by

$$\delta\omega = \int_{\omega_o - \pi\Delta f}^{\omega_o + \pi\Delta f} S_\omega(\omega)d\omega, \quad \text{and} \quad \Delta m \approx (2M_{eff}/\omega_o)\delta\omega \tag{1}$$

with $\Delta f$ the measurement bandwidth. If we assume for the roughness profile a single valued random function $h(r)$ of the in-plane position $r=(x,y)$ and a Gaussian height distribution [18], the rough area is given by $A_{rou}/A_{flat} = R_{rou} = \int_0^{+\infty} du \left(\sqrt{1+\rho^2 u}\right) e^{-u}$ [19] with $\rho = \sqrt{\langle(\nabla h)^2\rangle}$ the average local surface slope or $\rho = (\int_{0 \leq q \leq Q_c} q^2 \langle|h(q)|^2\rangle d^2 q)^{1/2}$ [20], and $A_{flat} = 2w_b L$ the average flat surface area. $\langle|h(q)|^2\rangle$ is the roughness spectrum, and $Q_c = \pi/a_o$ with $a_o$ a lower lateral cut-off. In addition, by assuming $Q \gg 1$ and $\omega_o/Q \gg 2\pi\Delta f$, Eq. (1) yields for $\Delta m$

$$\Delta m = \Delta m_{in} \left\{1 + (Q_{in}/Q_{gas,f})R_{rou}\right\}^{1/2} \tag{2}$$

where $\Delta m_{in} \approx 2M_{eff}(E_{th}/E_C)^{1/2}(\Delta f/\omega_o Q_{in})^{1/2}$ is the limit to mass sensitivity for flat surfaces for only thermomechanical noise [5], and $Q_{gas,f} = M_{eff}\omega_o\sqrt{K_B T/m}(PA_{flat})^{-1}$.

Our calculations will be performed for random self-affine rough surfaces observed in a wide spectrum of surface engineering processes [18]. In this case $\langle|h(q)|^2\rangle$ scales as $<|h(q)|^2> \propto q^{-2-2H}$ if $q\xi>>1$, and $<|h(q)|^2> \propto const$ if $q\xi<<1$ [18, 20, 21]. This is satisfied by the analytic model [21] $<|h(q)|^2>=(2\pi w^2\xi^2)/(1+aq^2\xi^2)^{(1+H)}$ with $a=(1/2H)[1-(1+aQ_c^2\xi^2)^{-H}]$ if $0<H<1$, and $a=1/2\ln(1+aQ_c^2\xi^2)$ if $H=0$. Small values of $H$ (~0) characterize jagged or irregular surfaces; while large values of $H$ (~1) surfaces with smooth hills-valleys (see inset in Fig. 1) [17, 20]. In addition, we obtain for the local slope the analytic expression $\rho=(w/\sqrt{2}\xi a)\{(1-H)^{-1}[(1+aQ_c^2\xi^2)^{1-H}-1]-2a\}^{1/2}$ [19], which further facilitates calculations of $\Delta m$. For other roughness models see ref. [21].

Figure 1 shows calculations of $\Delta m$ as a function of $Q_{in}/Q_{gas,f}$ for various roughness exponents $H$. Our calculations were performed for roughness amplitudes observed in real nanoresonator systems [11], and $a_o=0.3$ nm. As it is indicated with decreasing quality ratio $Q_{in}/Q_{gas,f}$ (or increasing gas dissipation), the limit to mass sensitivity becomes more sensitive to roughness changes at short length scales as the top most curve indicates for $Q_{in}>Q_{gas,f}$. The later is also directly shown in Fig.2 for various lateral correlation lengths $\xi$ and roughness amplitudes $w$. In the opposite limit for $Q_{in}<Q_{gas,f}$ (where the roughness influence is weak, by considering the asymptotic expansion $(1+y)^{1/2} \approx 1+y/2+....$ and weak local slopes ($\rho<<1$) so that $R_{rou}\cong 1+\rho^2/2$, we obtain the analytic form for the mass sensitivity $\Delta m \approx \Delta m_{in}\{1+(Q_{in}/2Q_{gas,f})+(Q_{in}/4Q_{gas,f})\rho^2+...\}$.

If we compare Figs. 1 and 2 we can infer that the influence of the roughness exponent $H$ plays significant role on the limit to mass sensitivity besides that of the most commonly used roughness parameters $w$ and $\xi$. In order, to gain better insight of its effect, we plot in Fig. 3 its direct influence on $\Delta m$ plotted for different roughness ratios $w/\xi$. It is shown that at small roughness exponents ($H\sim0$) the influence of the ratio $w/\xi$ is diminished. However, its influence is more distinct in the intermediate range of exponents $0.3<H<0.8$, which is the regime commonly observed in experimental systems [18].

Finally, we will consider in briefly also the contribution from other noise sources. Indeed, the limit to mass sensitivity due to temperature fluctuations $\Delta m_{Tem-Flu}$ is smaller than that of thermomechanical noise and can be neglected. On the other hand, the limit to mass sensitivity due to adsorption-desorption noise is influenced by morphology as $\Delta m_{a-d} \approx \Delta m_{a-d,flat} \{R_{rou}\}^{1/2}$ [23], which increases with increasing roughening. $\Delta m_{a-d,flat}$ is the mass sensitivity for flat surfaces. Therefore, $\Delta m_{a-d}$ can play role for the total mass sensitivity if morphology variations are under consideration. If we combine with Eq. (2) the total limit to mass sensitivity reads of the form $\Delta m = \Delta m_{in} \{1+(Q_{in}/Q_{gas,f})R_{rou}\}^{1/2} + \Delta m_{a-d,flat} \{R_{rou}\}^{1/2}$. Figure 4 shows calculations for this case for various ratios $\Delta m_{in}/\Delta m_{a-d,flat}$. It becomes clear that morphology effects from adsorption-desorption become significant if and only if $\Delta m_{in} \leq \Delta m_{a-d,flat}$.

In conclusion, we investigated at a first stage the simultaneous influence of thermomechanical and momentum exchange noise on the limit to mass sensitivity for nanoresonators. With increasing surface roughness, the limit to mass sensitivity increases significantly if the quality factor due to gas collisions is comparable or smaller than the intrinsic quality factor associated with thermomechanical noise. In addition, the influence of the roughness ratio *w/ξ* on the mass sensitivity becomes more distinct in the intermediate range of exponents *0.3<H<0.8* that are commonly observed in experiments. Notably, the morphology influence can be further enhanced if also mass loading due to adsorption-desorption noise plays significant role under specific conditions. In any case, our results indicate that the surface morphology could play important role on mass sensing of nanoresonators, which can be minimized by fabrication processes yielding smoother morphology and/or intrinsic quality factors comparable or larger to that imposed by the surrounding gas.

**Figure Captions**

**Figure 1** $\Delta m / \Delta m_{in}$ as a function of $Q_{in}/Q_{gas,f}$ for different roughness exponents $H$ as indicated, $w=3$ $nm$, and $\xi=60$ $nm$. The inset schematic is showing the influence of the roughness exponent $H$ for three surfaces with the same $w$ and $\xi$.

**Figure 2** $\Delta m / \Delta m_{in}$ as a function of $Q_{in}/Q_{gas,f}$ for different roughness amplitudes $w$ as indicated, $\xi=60$ $mm$, and $H=0.5$. The inset shows $\Delta m / \Delta m_{in}$ as a function of $Q_{in}/Q_{gas,f}$ for different correlation lengths $\xi$ as indicated, $H=0.5$, and $w=3$ $nm$.

**Figure 3** $\Delta m / \Delta m_{in}$ as a function of $H$ for $Q_{in}/Q_{gas,f}=100$ for different roughness ratios $w/\xi$ as indicated and $w=3$ $nm$.

**Figure 4** $\Delta m / \Delta m_{in}$ as a function of $H$ for $Q_{in}/Q_{gas,f}=100$ with $w/\xi=0.05$ and three different ratios of $\Delta m_{in}/\Delta m_{a-d,flat}$ as indicated with $w=3$ $nm$.

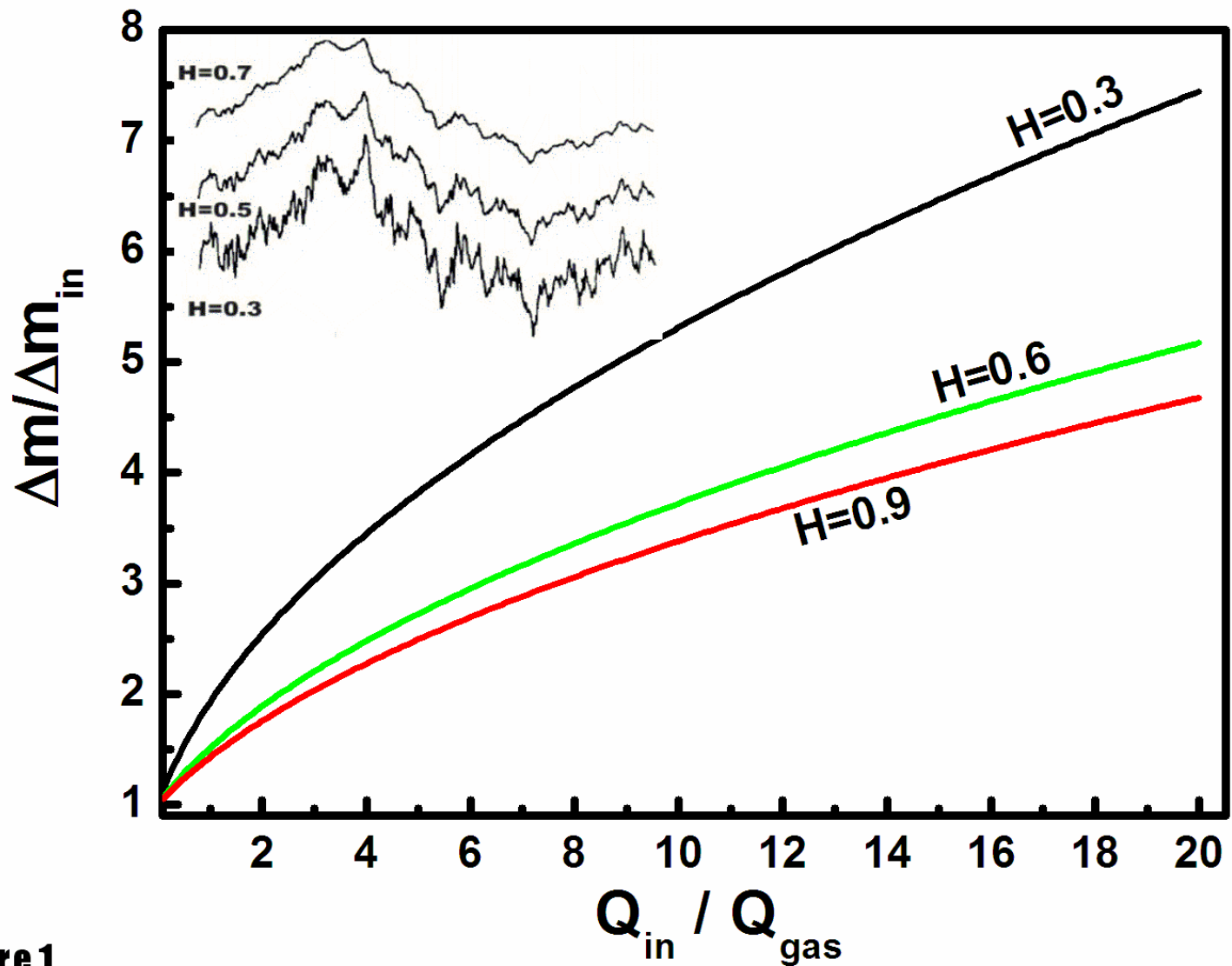

Figure 1

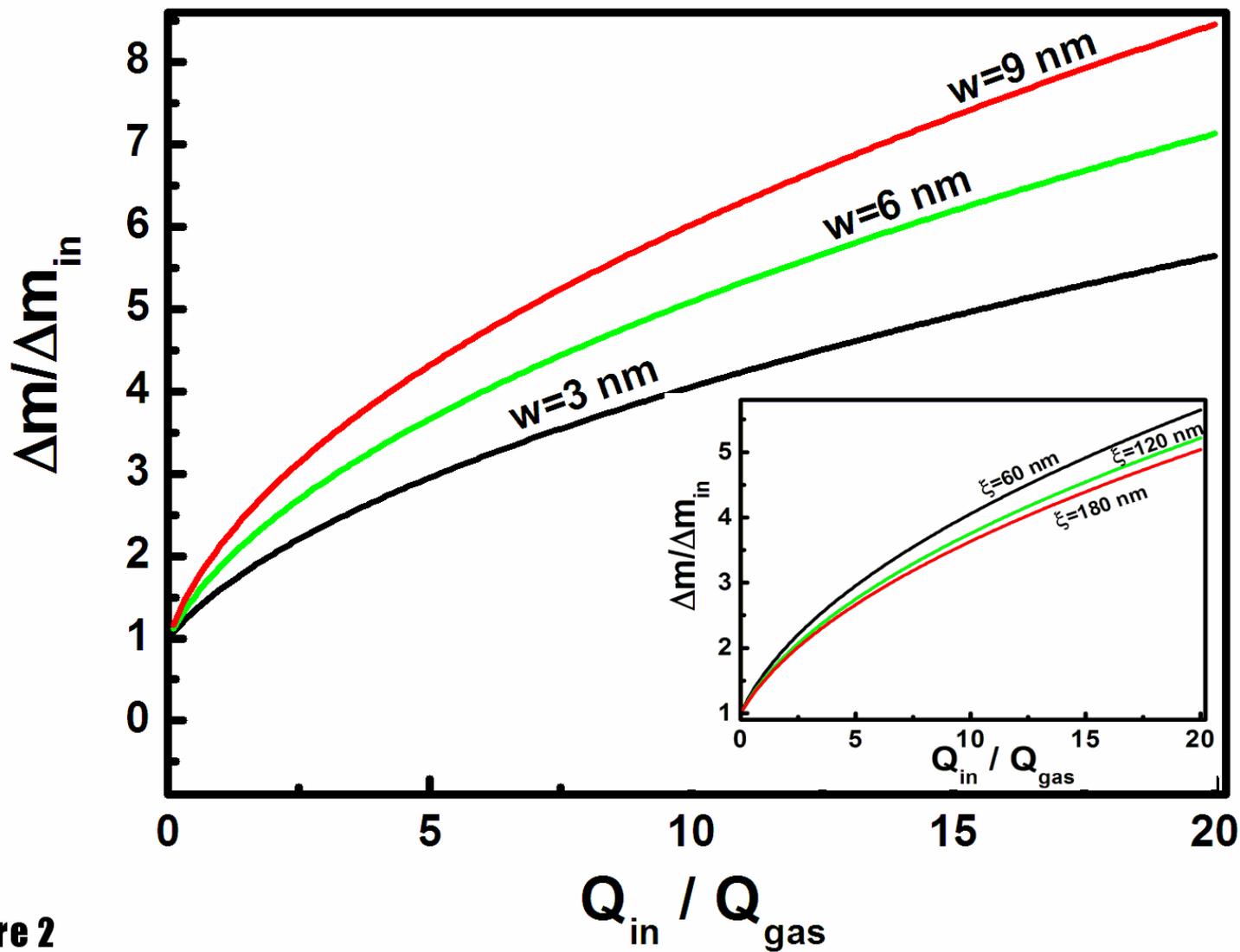

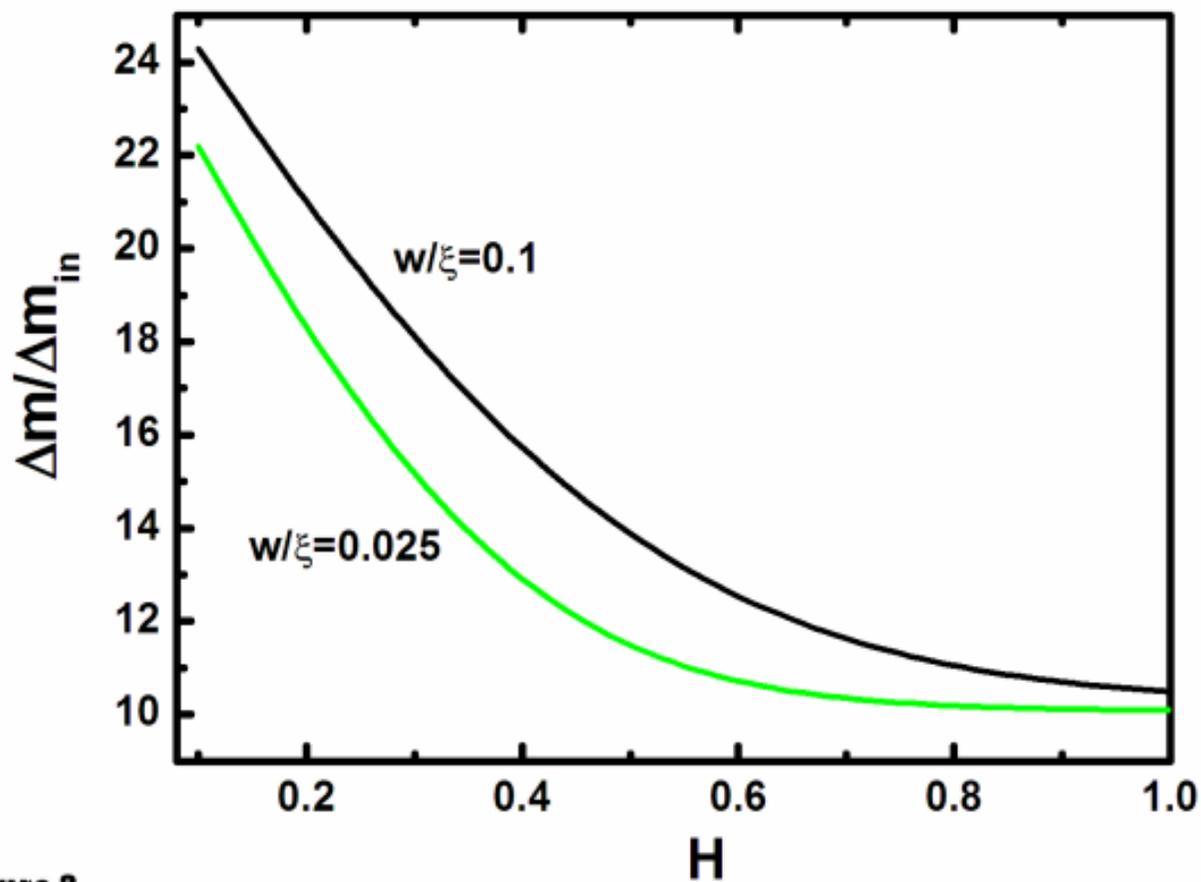

Figure 3

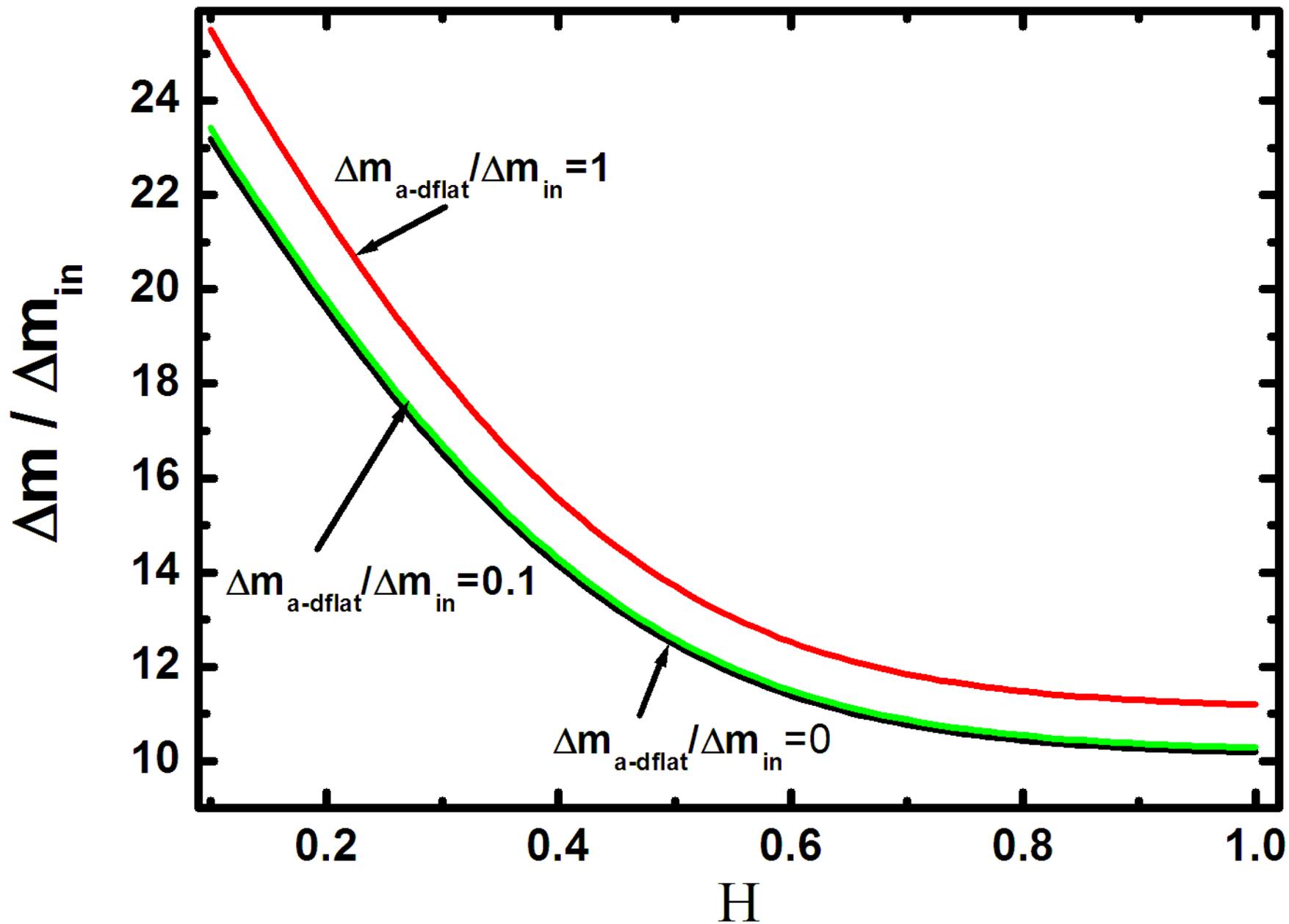

**Figure 4**